\documentclass[12pt]{article}
\usepackage{amsmath,amssymb}
\usepackage{amsfonts}
\usepackage[mathscr]{eucal}

\newcommand{\nn}{\nonumber \\}
\newcommand{\p}[1]{(\ref{#1})}
\newcommand{\be}{\begin{equation}}
\newcommand{\ee}{\end{equation}}
\newcommand{\bea}{\begin{eqnarray}}
\newcommand{\eea}{\end{eqnarray}}
\newcommand{\lb}{\label}
\begin{document}

\begin{center}
{\Large\bf New Super Calogero Models and
\vspace{0.2cm}

OSp(4$|$2) Superconformal
Mechanics} \footnote{Talk presented by E. Ivanov at the XIII International Conference ``Symmetry Methods in Physics'',
Dubna, July 6-9, 2009.}
\vspace{0.6cm}

{\large\bf
Sergey Fedoruk, Evgeny Ivanov,}
\vspace{0.2cm}

{\it Bogoliubov Laboratory of Theoretical Physics,
JINR, \\
141980, Dubna, Moscow Region, Russia}\\
{\tt fedoruk,eivanov@theor.jinr.ru}\\[8pt]
\vspace{0.3cm}

{\large\bf Olaf Lechtenfeld}
\vspace{0.2cm}

{\it Institut f{\" u}r Theoretische Physik, Leibniz Universit{\" a}t Hannover, \\ Appelstra{\ss}e 2,
D-30167 Hannover, Germany}\\
{\tt lechtenf@itp.uni-hannover.de}\\[8pt]
\end{center}
\vspace{0.5cm}

\begin{abstract}
\noindent
We report on the new approach to constructing superconformal extensions of the Calogero-type systems with
an arbitrary number of involved particles. It is based upon the superfield gauging of non-abelian isometries
of some supersymmetric matrix models. Among its applications,
we focus on the new ${\cal N}{=}4$ superconformal system yielding the U(2) spin Calogero model in the bosonic sector, and the one-particle case of
this system, which is a new OSp(4$|$2) superconformal mechanics with non-dynamical U(2) spin variables.
The characteristic feature of these models is that the strength of the conformal inverse-square potential
is quantized.
\end{abstract}

\section{Motivations and contents}
The conformal Calogero model \cite{C} describes $n$
identical particles interacting pairwise through an inverse-square
potential
\be
V_C=\sum_{a\neq b} \frac{g}{(x_a - x_b)^2}\,, \qquad a,b=1,...,n\,. \lb{CalPot}
\ee
%\item
It is a nice example of integrable $d=1$ system. This simplest ($A_{n-1}$) Calogero model has
some integrable generalizations, both conformal and non-conformal \cite{C,C2}.

As for superconformal  extensions of the Calogero models (s-C models in what follows),
the basic facts about them can be shortly summarized as follows:

\begin{itemize}
\item ${\cal N}=2$ superextension of the model \p{CalPot} and its some generalizations for any $n$
was given by Freedman and Mende in 1990 \cite{FM} (see also \cite{Vas} for ${\cal N}=1$ extensions).

\item First attempts toward ${\cal N}=4$ extensions were undertaken by Wyllard in 2000 \cite{W}. Further progress
was achieved in refs. \cite{BGK} - \cite{KLP}.

\item Until recently, ${\cal N}{=}4$ s-C models generalizing \p{CalPot} for a generic $n$ were not constructed.
\end{itemize}

At the same time, s-C systems are of great interest from various points of view. In 1999, Gibbons and Townsend \cite{GT}
suggested that ${\cal N}=4$ s-C models might provide a microscopic description of the extreme Reissner-Nordstr\"{o}m black hole
in the near-horizon limit and, even more, be one of the faces of the hypothetical M-theory.
Also, this sort of models can bear a tight relation to AdS/CFT and brane stuff (M-theory, strings, etc), quantum Hall effect
(see, e.g., \cite{C2}, \cite{Poly0} - \cite{Poly1}), etc. One-particle prototype of s-C systems is the superconformal mechanics.
The first ${\cal N}=2$ and ${\cal N}=4$
variants of the latter were constructed and studied by Akulov and Pashnev in 1983 \cite{AP},
Fubini and Rabinovici in 1984 \cite{FR} and  Ivanov et al in 1989 \cite{IKL}. These models attract a lot of attention
mainly because of their intimate relationships to the description of the black-hole type
solutions of supergravity (see e.g. \cite{BH,MS}).

Recently, we suggested a universal approach to  s-C models for an arbitrary number $n$
of interacting particles, including the ${\cal N}{=}4$ models \cite{FIL1}.
It is based on the superfield gauging  of non-abelian isometries of some supersymmetric matrix
models along the line of ref. \cite{DI}.\\

\noindent{\it This new approach is based upon the following two primary principles}:
\begin{itemize}
\item ${\rm U}({n})$ gauge invariance for ${n}$-particle s-C models;
\item ${\cal N}$-extended superconformal symmetry.
\end{itemize}
%\item

\noindent{\it The models constructed in this way display the following salient features}:
\begin{itemize}
\item Their bosonic sector is:\\
- the standard $A_{n-1}$  Calogero model for ${\cal N}{=}1$ and ${\cal N}{=}2$ cases, \\
- a new variant of the ${\rm U}(2)$-spin Calogero
model \cite{C2} in the ${\cal N}{=}4$ case;
\item In the ${\cal N}{=}2$ case there arise new  superconformal extensions (different from those of Freedman and Mende);
\item In the ${\cal N}{=}4$ case the gauge approach directly yields ${\rm OSp}(4|2)$
as the superconformal group, but the general ${\cal N}{=}4, d{=}1$ superconformal group
$D(2,1;\alpha)$ can be incorporated as well;
\item
The center-of-mass coordinate in the ${\cal N}{=}4$ case is not decoupled, and it acquires
a conformal potential on shell. So a new model of ${\cal N}{=}4$ superconformal
mechanics emerges in the $n{=}1$ limit \cite{FIL2,FIL3} (see also \cite{KL}).
\end{itemize}

In the present talk we give a brief account of this gauge approach, with the main focus on the ${\cal N}=4$ super Calogero model and
the new ${\cal N}{=}4$ superconformal mechanics just mentioned.

\section{Bosonic Calogero as a gauge matrix model}

The nice interpretation of the model \p{CalPot} as a gauge model was given in \cite{Poly0,Gorsky}.

The starting point of this approach is the ${\rm U}(n)$, $d{=}1$ matrix gauge theory which involves:\\

- an hermitian $n{\times}n$-matrix field $X_a^b(t)$, $(\overline{X_a^b} =X_b^a)$, $(a,b=1,\ldots ,n)\,$; \\

- a complex ${\rm U}(n)$-spinor field $Z_a(t)$, $\bar Z^a = \overline{(Z_a)}\,$;  \\

- $n^2$ non--propagating ${\rm U}(n)$ ``gauge fields'' $A_a^b(t)$, $\overline{(A_a^b)} =A_b^a\,$.\\

The invariant action is written as \cite{Poly0}:
\be
S_0 = \int dt  \,\Big[\, {\rm Tr}\left(\nabla\! X \nabla\! X \right) +
{\textstyle\frac{i}{2}} (\bar Z \nabla\! Z - \nabla\! \bar Z Z) + c\,{\rm Tr} A  \,\Big],\lb{ActB}
\ee
where
$$
\nabla\! X = \dot X +i [A,X], \quad \nabla\! Z = \dot Z + iAZ\;.
$$
It respects the following invariances:
\begin{itemize}
\item The $d=1$ conformal ${\rm SO}(1,2)$ invariance realized by the transformations:
$$
\delta t = a\,, \qquad \partial_t^3 {a} = 0 \,,
$$
$$
\delta X_a^b = {\textstyle\frac{1}{2}}\, \dot{a} X_a^b,\qquad \delta Z_a = 0,\qquad \delta A_a^b
= -\dot{a}A_a^b\,.
$$
\item The invariance under the local ${\rm U}(n)$ transformations:
$$
X \!\rightarrow  g X g^\dagger , \quad  Z \!\rightarrow  g Z , \quad A \!\rightarrow  g A g^\dagger +i
\dot g g^\dagger\,,
$$
where $g(\tau )\in {\rm U}(n)\,$.

\end{itemize}

Using this gauge ${\rm U}(n)$ freedom, one can impose the following gauge conditions:
\be
X_a^b = x_a \delta_a^b , \qquad  \bar Z^a = Z_a\,. \lb{GF1}
\ee
As the next step, one makes use of the algebraic equations of motion
$$
\delta A^a_a: \; (Z_a)^2 =c,  \quad \delta A^a_b\;\; (for \; a\neq b): \;
A_a^b = \frac{Z_a Z_b}{2(x_a - x_b)^2}.
%\,\, {\scriptstyle a\neq b}
$$
Substituting the expression for $A^a_b$ back into the gauge-fixed form of the action \p{ActB},
one recovers the standard Calogero action
$$
S_0 = \int dt  \,\Big[\, \sum_{a} \dot x_a \dot x_a - \sum_{a\neq b} \frac{c^2}{4(x_a -
x_b)^2}\,\Big].
$$

Our approach to supersymmetric extensions of the Calogero models is just supersymmetrization of the above
gauge approach, with the fields $X, Z, A\,$ substituted by the appropriate $d=1$ superfields.

\section{${\cal N}=1$ superconformal Calogero system}

We start with a brief account of the ${\cal N}{=}1, d{=}1$ supersymmetric version.

The point of departure in this case is the one-dimensional ${\cal N}=1$ supersymmetric ${\rm U}(n)$ gauge theory which involves:\\

- an even matrix superfield $\mathscr{X}_a^b(t, \theta)$, $(\mathscr{X})^\dagger =\mathscr{X}\,$, \\

- an even ${\rm U}(n)$-spinor superfield $\mathcal{Z}_a (t, \theta)$,
$\bar{\mathcal{Z}}{}^a (t, \theta) = (\mathcal{Z}_a)^\dagger $,  \\

- an odd gauge connections $\mathscr{A}_a^b(t, \theta )$, $(\mathscr{A})^\dagger = -\mathscr{A}$.\\
%\item

The invariant action is written as an integral over the ${\cal N}{=}1, d{=}1$ superspace:
\be
S_{1}\! =\! -i\!\int\! dt d\theta \Big[ {\rm Tr} \left( \nabla_t \mathscr{X} \mathscr{D}
\mathscr{X} + c \mathscr{A} \right) + {\textstyle\frac{i}{2}}(\bar {\mathcal{Z}} \mathscr{D}
\mathcal{Z} - \mathscr{D}\bar {\mathcal{Z}} \mathcal{Z} )\Big], \lb{ActN1}
\ee
where
$$
\mathscr{D} \mathscr{X} =  D \mathscr{X} +i [ \mathscr{A} , \mathscr{X}], \quad \nabla_t
\mathscr{X} = -i\mathscr{D}\mathscr{D} \mathscr{X}, \quad \mathscr{D} \mathcal{Z} =  D
\mathcal{Z} +i \mathscr{A} \mathcal{Z},
$$
$$
D = \partial_{\theta} +i\theta\partial_{t}\,, \quad \quad \{D, D \} = 2i\,
\partial_{t}\,.
$$

The action \p{ActN1} possesses ${\cal N}{=}1$ superconformal ${\rm OSp}(1|2)$ invariance:
%\begin{center}
$$
\delta t= -i\,\eta\theta t, \qquad \delta \theta = \eta t,
$$
$$
\delta \mathscr{X}= -i\,\eta\theta\,\mathscr{X},\quad \delta \mathscr{A} =
i\,\eta\theta\,\mathscr{A},\quad \delta \mathcal{Z} = 0\,,
$$
and gauge ${\rm U}(n)$ invariance:
$$
\mathscr{X}^{\,\prime} =
e^{i\tau} \mathscr{X} e^{-i\tau}, \quad \mathcal{Z}^{\prime} = e^{i\tau} \mathcal{Z}, \quad
\mathscr{A}^{\,\prime} =  e^{i\tau} \mathscr{A} e^{-i\tau} - i e^{i\tau} D e^{-i\tau},
$$
where $\tau_a^b(t, \theta) \in u(n)$ is an hermitian matrix parameter.

One can choose WZ gauge for the spinor connection:
\be
\mathscr{A} = i\theta A(t)\,. \lb{WZ1}
\ee
After integrating  over $\theta$ s in the gauge-fixed form of \p{ActN1}
and eliminating auxiliary fields, one obtains:
\be
S_{1} = S_{0}+S^\Psi_{1},\qquad S^\Psi_{1}=-i\,{\rm Tr}\int dt \,\Psi \nabla \Psi\,, \lb{ActN12}
\ee
where $\Psi = -iD\mathscr{X}|$ and $\nabla \Psi = \dot \Psi +i [A,\Psi]\,$.
The bosonic part $S_0$ of $S_1$ in \p{ActN12} is just the ``gauge-unfixed'' Calogero action \p{ActB}.

After integrating out non-propagating gauge fields $A^a_b$ from the total action $S_1$ and fixing the residual
U($n$) gauge freedom of the WZ gauge \p{WZ1} in the same way as in \p{GF1},
we obtain an ${\cal N}{=}1$ superconformal action which contains $n$ bosonic fields $x_a$
with the standard conformal Calogero potential \p{CalPot} accompanied by interactions with
$n^2$ physical fermionic fields $\psi^a_b$.
%\end{itemize}
%\end{frame}

\section{${\cal N}=4$ superconformal Calogero}

${\cal N}{=}4, d{=}1$ models are naturally formulated in the $d=1$ version of harmonic superspace (HSS) \cite{IL} :
$$
(t,\theta_i, \bar\theta^k, u_i^\pm),\qquad {i,k =1,2}.
$$
%It is a reduction of ${\cal N}{=}2, d{=}4$ HSS \cite{HSS}.
Bosonic ${\rm SU}(2)$-doublets $u_i^\pm$
are harmonic coordinates, with the basic relation $u^{+i}u_i^-=1\,$.
The main feature of HSS is the presence of harmonic analytic subspace in it (an analog of chiral superspace),
closed under the action of ${\cal N}{=}4$ supersymmetry:
$$
(\zeta,u)=(t_A,\theta^+,
\bar\theta^+, u_i^\pm),
$$
$$
\quad t_A=t-i(\theta^+ \bar\theta^- +\theta^-\bar\theta^+),\quad
\theta^\pm=\theta^i u_i^\pm,\;\bar\theta^\pm=\bar\theta^i u_i^\pm\,.
$$
The integration measures in the full HSS and its analytic subspace are defined, respectively, as:
$$
\mu_H =dudtd^4\theta,\qquad \mu^{(-2)}_A=dud\zeta^{(-2)}.
$$

The ${\cal N}=4, d=1$ supergauge theory which generalizes the bosonic and ${\cal N}{=}1$
examples described above involves the following superfields:
\vspace{0.3cm}

\begin{itemize}
\item  The hermitian matrix superfields $\mathscr{X}=(\mathscr{X}_a^b)$ (multiplets ({\bf 1,4,3})):
$$
\mathscr{D}^{++} \,\mathscr{X}=0, \qquad
\mathscr{D}^{+}\mathscr{D}^{-} \,\mathscr{X}=0, \quad
 (\mathscr{D}^{+}\bar{\mathscr{D}}^{-} +\bar{\mathscr{D}}^{+}\mathscr{D}^{-})\, \mathscr{X}=0
;
$$
\item  Analytic superfields $\mathcal{Z}^+_a(\zeta,u)$ (multiplets ({\bf 4,4,0})):
$$
\mathscr{D}^{++} \mathcal{Z}^+=0, \quad {D}^{+} \mathcal{Z}^+ =0\,,\quad
 \bar{D}^{+} \mathcal{Z}^+ =0\,;
$$
\item  The analytic gauge matrix connection $V^{++}(\zeta,u)\,.$
It specifies the gauge-covariant derivatives (harmonic and spinor):
$$
\mathscr{D}^{++}\mathcal{Z}^+ = (D^{++} + i\,V^{++})\mathcal{Z}^+,\quad \mathscr{D}^{++} \mathscr{X} = D^{++}
\mathscr{X} + i\,[V^{++} ,\mathscr{X}]\,, \;{\rm etc}.
$$
\end{itemize}

The ${\cal N}{=}4$ superconformal action is a sum of three terms:
%the action
\be
S_4 = S_{\mathscr{X}} + S_{WZ} +  S_{FI}\,,\lb{ActN4}
\ee
where
\be
S_{\mathscr{X}} =-{\textstyle\frac{1}{2}}\int \mu_H  {\rm Tr} ( \mathscr{X}^{\,2}
), \qquad\qquad S_{WZ} = {\textstyle\frac{1}{2}}\int \mu^{(-2)}_A  \mathcal{V}_0
\widetilde{\,\mathcal{Z}}{}^+ \mathcal{Z}^+, \lb{ActXWZ}
\ee
\be
S_{FI} ={\textstyle\frac{i}{2}}\,c\int \mu^{(-2)}_A  \,{\rm Tr} \,V^{++} \lb{FI}
\ee
and $\mathcal{V}_0(\zeta,u)$ is a real analytic superfield, which is related to
$\mathscr{X}_0\equiv {\rm Tr}\left( \mathscr{X} \right)$ by the integral
transform
\be
\mathscr{X}_0(t,\theta_i,\bar\theta^i) = \int du\, \mathcal{V}_0 \left(t_A, \theta^+,
\bar\theta^+, u^\pm \right) \Big|_{\theta^\pm=\theta^i u^\pm_i,\,\,\,
\bar\theta^\pm=\bar\theta^i u^\pm_i}\,,
\ee
$$
\mathcal{V}_0{}' = \mathcal{V}_0 + D^{++}\Lambda^{--}\,.
$$

%\frametitle{Invariances of the action}
The action \p{ActN4} respects the following set of invariances:
\begin{itemize}
\item ${\cal N}{=}4$ superconformal invariance under the supergroup
$D(2,1;\alpha=-1/2) \simeq {\rm OSp}(4|2)$:
$$
\delta \mathscr{X}= -\Lambda_0\,\mathscr{X},\quad \delta \mathcal{Z}^+ =
\Lambda\,\mathcal{Z}^+,\quad \delta V^{++} = 0, \quad \delta\mathcal{V}_0 = -2\Lambda \mathcal{V}_0,
$$
\begin{center}
$
\Lambda = 2i\alpha(\bar\eta^-\theta^+ -
\eta^-\bar\theta^+)$,\quad
$\Lambda_0 = 2\Lambda- D^{--} D^{++}\Lambda\,$;
\end{center}
\item Gauge ${\rm U}(n)$ invariance:
$$
\mathscr{X}^{\,\prime} =  e^{i\lambda} \mathscr{X} e^{-i\lambda} , \qquad
\mathcal{Z}^+{}^{\prime}
= e^{i\lambda} \mathcal{Z}^+,
$$
$$
V^{++}{}^{\,\prime} =  e^{i\lambda}\, V^{++}\, e^{-i\lambda} - i\, e^{i\lambda} (D^{++}
e^{-i\lambda})\, ,
$$
where $\lambda_a^b(\zeta, u^\pm) \in u(n) $ is the `hermitian' analytic matrix
parameter, $\widetilde{\lambda} =\lambda\,$.
\end{itemize}

Like in the ${\cal N}{=}1$ case, using the ${\rm U}(n)$ gauge freedom we can choose
the WZ gauge for $V^{++}$:
\be
V^{++} =-2i\,\theta^{+}
 \bar\theta^{+}A(t_A). \lb{WZ2}
\ee
In this gauge:
\bea
S_4 &{=}& S_b + S_f\,, \lb{ActN4c} \\
%\begin{eqnarray} \nonumber
S_b &{=}& \int\!\! dt \,\Big[{\rm Tr} \left( \nabla X\nabla X +c \,A \right)
+ {\textstyle\frac{i}{2}}X_0\! \left(\bar Z_k \nabla Z^k \!\! - \!\!\nabla \bar Z_k Z^k\right)
+ {\textstyle\frac{n}{8}}(\bar Z^{(i} Z^{k)})(\bar Z_{i} Z_{k})
\!\Big], \nonumber
\\
S_f &{=}& -i{\rm Tr} \int dt \left( \bar\Psi_k \nabla\Psi^k
-\nabla\bar\Psi_k \Psi^k
\right) -\int dt  \,\frac{\Psi^{(i}_0\bar\Psi^{k)}_0 (\bar Z_{i}
Z_{k})}{X_0}\,. \nonumber
\eea
Here $\mathscr{X}= X(t_A) + \theta^- \Psi^{i}(t_A) u_i^+ + \bar\theta^- \bar\Psi^{i}(t_A)u_i^+ + \ldots\,$,
$X_0 \equiv {\rm Tr} (X), \quad\Psi_0^i \equiv {\rm Tr} (\Psi^i), \quad\bar\Psi_0^i \equiv
{\rm Tr} (\bar\Psi^i)\,,$ $\mathcal{Z}^+ = Z^{i}(t_A)u_i^+ + \ldots\,$.\\

Let us study the bosonic limit of the action \p{ActN4c}. To pass to this limit, one needs:
\begin{itemize}
\item to impose the gauge $X_a^b =0$, $a\neq b$;
\item to eliminate $A_a^b$, $a\neq b$, by their algebraic equations of motion;
\item to pass to the new fields $Z^\prime{}^i_a =
(X_0)^{1/2}Z^i_a$ (in what follows, we shall omit primes).
\end{itemize}

As a result, we obtain the following bosonic action
\be
S_{b} = \!\!\int\!\! dt \Big\{\! \sum_{a} \dot x_a \dot x_a +
{\textstyle\frac{i}{2}}\sum_{a} (\bar Z_k^a \dot Z^k_a - \dot {\bar Z}{}_k^a Z^k_a)  +
\sum_{a\neq b} \frac{{\rm Tr}(S_a S_b)}{4(x_a - x_b)^2} - \frac{n\,{\rm
Tr}(\hat S \hat S)}{2(X_0)^2}\!\Big\}, \lb{BosN4}
\ee
where
\be
(S_a)_i{}^j \equiv \bar Z^a_i Z_a^j, \quad
(\hat S)_i{}^j \equiv \sum_a \left[ (S_a)_i{}^j -
{\textstyle\frac{1}{2}}\delta_i^j(S_a)_k{}^k\right]. \lb{Sdef}
\ee
The fields $Z^k_a$ are subject to the constraints
\be
\bar Z_i^a Z^i_a =c \qquad \forall \, a \,.\lb{ConstrZ}
\ee

Since the fields $Z_k^a, {\bar Z}{}_k^a$ are described by the Lagrangian of the first order in the
time derivative, we are led quantize them by the Dirac quantization method:
$$
{\textstyle\frac{i}{2}}\int dt
\sum_{a} (\bar Z_k^a \dot Z^k_a - \dot {\bar Z}{}_k^a Z^k_a) \qquad
\Rightarrow\qquad
[\bar Z^a_i, Z_b^j]_{{}_D}= i\delta^a_b\delta_i^j.
$$
Now it is easy to check that the quantities $S_a$ defined in \p{Sdef} form, for each ${a}$, $u(2)$
algebras
$$
[(S_a)_i{}^j, (S_b)_k{}^l]_{{}_D}= i\delta_{ab}\left\{\delta_i^l(S_a)_k{}^j-
\delta_k^j(S_a)_i{}^l \right\}.
$$

Modulo center-of-mass conformal potential, the bosonic action \p{BosN4} can be written as
\be
S^\prime_{b} = \int dt \Big\{\! \sum_{a} \dot x_a \dot x_a +
\sum_{a\neq b} \frac{{\rm Tr}(S_a S_b)}{4(x_a - x_b)^2} \Big\}.\lb{BosN4a}
\ee
It is none other than the action of integrable ${\rm U}(2)$-spin Calogero
model \cite{C2}.

\section{OSp(4$|$2) superconformal mechanics}
The ${n=1}$ case of the ${\cal N}{=}4$ Calogero model, as distinct from the ${\cal N}{=}1$
and ${\cal N}{=}2$ models, already yields a conformal inverse-square potential at the component level (for the center-of-mass
coordinate) and so yields a non-trivial ${\cal N}{=}4$ superconformal mechanics model \cite{FIL2}.
%\item

The corresponding superfield action is
\be
S =S_{\mathscr{X}} + S_{FI} + S_{WZ}\,, \lb{ActSC}
\ee
where
\bea
&& S_{\mathscr{X}} =-{\textstyle\frac{1}{2}}\int \mu_H \, \mathscr{X}^{\,2}\,, \quad
S_{FI} ={\textstyle\frac{i}{2}}\,\,c\int \mu^{(-2)}_A \,V^{++}\,, \quad
S_{WZ} = {\textstyle\frac{1}{2}}\,\int \mu^{(-2)}_A \, \mathcal{V}\,
\bar{\mathcal{Z}}{}^+\, {\mathcal{Z}}^+\, ,\nn
&& D^{++} \,\mathscr{X}=0\,, \qquad D^{+}{D}^{-} \,\mathscr{X}=
    \bar D^{+}\bar D^{-}\, \mathscr{X}= (D^{+}\bar D^{-} +\bar D^{+}D^{-})\, \mathscr{X}=0\,, \nn
&& \mathscr{D}^{++} \,{\mathcal{Z}}^+\equiv (D^{++} + i\,V^{++})
\,{\mathcal{Z}}^+=0\,, \qquad
\mathscr{D}^{++}\,\bar{\mathcal{Z}}{}^+\equiv (D^{++} - i\,V^{++}) \,\bar{\mathcal{Z}}{}^+=0\,, \nn
&& \delta V^{++}= -D^{++}\lambda\,, \qquad
\delta{\mathcal{Z}}^+ = i\lambda {\mathcal{Z}}^+\,, \qquad \delta \mathcal{V} = D^{++}\lambda^{--}\,.
\eea

The actions $S_{FI}$ and $S_{WZ}$ are invariant under the general ${\cal N}{=}4$ superconformal
group $D(2,1;\alpha)$ for arbitrary $\alpha$, while  $S_{\mathscr{X}}$ - only with respect to
$D(2,1;\alpha = -1/2) \sim OSp(4|2)$, so the total action is $OSp(4|2)$-superconformal.

The action \p{ActSC} can be generalized to any $\alpha$, but at cost of nonlinear action for $\mathscr{X}$.
So we limit our presentation here to the case of $\alpha = -1/2\,$.

The off-shell component content of the model is $({\bf 1, 4, 3}) \oplus  ({\bf 4, 4, 0}) \oplus V^{++}
= ({\bf 1, 4, 3}) \oplus ({\bf 3, 4, 1})\,$. So it is reducible.
%\item
The on-shell content can be most clearly revealed in the WZ gauge $V^{++} = -2i\theta^+\bar\theta^+ A(t)$.
After eliminating auxiliary fields from the total action, the latter takes the form
\be
S = S_b+ S_f\,,
\ee
where
\bea
&& S_b =  \int dt \,\Big[\dot x\dot x  + {\textstyle\frac{i}{2}} \left(\bar
z_k \dot
z^k -
\dot{\bar z}_k z^k\right)-\frac{(\bar z_k
z^{k})^2}{16x^2} -A \left(\bar z_k z^{k} -c \right) \Big] \,, \nn
&& S_f =  -i\, \int dt \,\left( \bar\psi_k \dot\psi^k -\dot{\bar\psi}_k
\psi^k \right)-
    \int dt \, \frac{\psi^{i}\bar\psi^{k} z_{(i} \bar z_{k)}}{x^2}\,.
\eea
Varying with respect to ${A(t)}$ as a Lagrange multiplier yields the constraint
\be
\bar z_k z^{k}=c\,. \lb{ConstrSC}
\ee
After properly fixing the residual ${\rm U(1)}$ gauge invariance, one can solve this constraint in terms of
two independent fields $\gamma(t)$ and $\alpha(t)$ as
\be
z^{1} = \kappa \cos{\textstyle\frac{\gamma}{2}}\,e^{i\alpha/2}\,,\quad
z^{2}= \kappa \sin{\textstyle\frac{\gamma}{2}}\,e^{-i\alpha/2}\,,\quad
\kappa^2=c\,.
\ee
Then the bosonic action takes the form
\be
S_b = \int dt \,\Big[\dot x\dot x  -\frac{c^2}{16\,x^2} - \frac{c}{2}\,
\cos\gamma\, \dot \alpha  \Big] \,. \lb{BosActSC}
\ee
It is a sum of the standard conformal mechanics action for the variable $x(t)$,
and the $S^2$ Wess-Zumino term for the angular variables $\gamma(t)$ and $\alpha(t)$, of the first-order
in time derivative. Taken separately, this term provides an example of Chern-Simons mechanics \cite{ChS,ChS1}.
The variables $\gamma(t)$ and $\alpha(t)$ (or  $z^{k}$ and $\bar z_k$ in the manifestly SU(2)
covariant formulation) become spin degrees of freedom (``target harmonics'') upon quantization.

%\frametitle{Quantization}
We quantize by Dirac procedure,
\bea
&& [X, P] = i\,, \; [Z^i, \bar Z_j] = \delta^i_j \,, \; \{\Psi^i,
\bar\Psi_j\}=
-{\textstyle\frac{1}{2}}\,\delta^i_j \,, \nn
&& P = \frac{1}{i}\partial/\partial X, \; \bar Z_i=v^+_i, \,Z^i=  \partial/\partial v^+_i, \;
\Psi^i=\psi^i, \,\bar\Psi_i= -{\textstyle\frac{1}{2}}\,
\partial/\partial\psi^i\,.
\eea
The wave function is subject to the constraint
\be
D^0 \Phi=\bar Z_i Z^i \Phi=v^+_i\frac{\partial}{\partial
v^+_i}\,\Phi=c\,\Phi\,, \lb{ConstrQ}
\ee
whence
\be
\Phi=A^{(c)}_{1}+ \psi^i B^{(c)}_i +\psi^i\psi_i A^{(c)}_{2} \,,\lb{WF}
\ee
and the component fields are collections of the ${\rm SU(2)}$ irreps with the isospins ${c/2}$, ${(c+1)/2}$, ${(c-1)/2}$
and ${c/2}$, respectively (the component fields depend on $X\,$).

The constant ${c}$ gets quantized,  ${c\in \mathbb{Z}}\,$, as a consequence of requiring the wave function \p{WF}
with the constraint \p{ConstrQ} to be single--valued. The same phenomenon takes place in the general $n$-particle
${\cal N}{=}4$ Calogero system due to the constraints \p{ConstrZ} which, after quantization, become analogs
of \p{ConstrQ}. This is also in agreement with the analogous arguments in the topological Chern--Simons
quantum mechanics \cite{ChS1}, which are based upon the path integral quantization.\footnote{Actually, in the considered
gauge approach to Calogero-type models the constant $c$ is quantized in all cases, including the purely bosonic one, due to
its appearance as a strength of the term $c \int dt\, {\rm Tr} A$ in the total component actions \cite{Poly0}
(see also a recent paper \cite{IKS}).}

The quantum Hamiltonian is given by the expression
\be
\mathbf{H} =\frac{1}{4}\,\left(P^2  +\frac{l(l+1)}{X^2}\right), \quad l = (c/2, (c+1)/2, (c-1)/2, c/2)\,.
\ee

The basic distinguishing feature of the new superconformal mechanics model is that the bosonic sector of the space of its quantum
states (with fermionic states neglected) is a direct product of the space of states of the standard conformal mechanics (parametrized  by $X$)
and a fuzzy sphere \cite{Mad} (parametrized by $Z^i, \bar Z_i$). The full wave functions are irreps of SU(2). The whole space
of states (with fermions), shows no any product structure.

One can ask what is the brane analog of this new superconformal mechanics via AdS$_2$/CFT$_1$ correspondence. The preliminary
answer is that it is some superparticle evolving on AdS$_2$ and coupled to the external magnetic charge
via WZ term.

\section{Summary and outlook}

Let us summarize the results presented in the Talk and outline directions of the further studies.

\begin{itemize}
\item We proposed a new gauge approach to the construction
of superconformal Calogero-type systems. The characteristic features
of this approach are the presence of auxiliary supermultiplets
with WZ type actions, the built-in superconformal
invariance and the emergence of the Calogero coupling constant as a strength
of the FI term of the ${\rm U}(1)$ gauge (super)field. This strength is quantized.
\item We used the ${\rm U}(n)$ gauging and obtained
superextensions
of the $A_{n-1}$ Calogero model. Superextensions of other conformal Calogero models
could be obtained by choosing other gauge groups.
\item The ${\cal N}{=}4$ action presented is invariant under
$D(2,1;-1/2)\cong {\rm OSp}(4|2)$. It can be easily generalized to an arbitrary
$\alpha\,$.\footnote{For the particular case of $n=1$ (superconformal mechanics) such a generalization
was given in a recent paper \cite{FIL3}. For $\alpha\neq 0$, the general $n$ particle action
was given in \cite{FIL1}.}
\item We constructed a new ${\cal N}{=}4$ superconformal mechanics with the ${\rm OSp}(4|2)$ invariance as
the extreme ${n=1}$ case of our ${\cal N}{=}4$ Calogero system. After quantization it yields
a fuzzy sphere in the bosonic sector.
\end{itemize}

As the mainstream directions of the future work we would like to distinguish {\bf (i)} construction of the full quantum version of
the new ${\cal N}=4$ super Calogero model for any number of particles (as well as of the new ${\cal N}=1$ and ${\cal N}=2$ extensions);
{\bf (ii)} elucidating the relationships of this ${\cal N}=4$ super Calogero system to the ${\cal N}=4$ Calogero-type systems
considered in \cite{W} - \cite{KLP} and to the black-hole stuff; {\bf (iii)} analysis of possible integrability properties
of the new super Calogero systems (searching for their Lax pair representation, etc).

\section*{Acknowledgements}

\noindent
We acknowledge a support from a DFG grant, project No 436 RUS/113/669 (E.I. \& O.L.),
the RFBR grants 08-02-90490, 09-02-01209 and 09-01-93107 (S.F. \& E.I.) and a grant of the Heisenberg-Landau Program.

\end{document}